\documentclass[doublespacing]{elsart}

\usepackage{graphicx}
\usepackage{amsmath}
\usepackage{amssymb}
\usepackage{bm}
\usepackage{setspace}
\usepackage{color}
\usepackage{colortbl}

\providecommand{\Journal}[4] {#1 {\bf #2} (#4) #3}
\providecommand{\ARNPS}{Ann. Rev. Nucl. Part. Sci. } %
\providecommand{\APP}{Ann. Phys. -Paris } %
\providecommand{\EPJC}{Eur. Phys. J. C } %
\providecommand{\HEP}{High Energy Phys. Nucl. Phys. (in Chinese) }%
\providecommand{\LNC}{Lett. Nuovo Cim. } %
\providecommand{\MPLA}{Mod. Phys. Lett. A } %
\providecommand{\NCA}{Nuovo Cim. A } %
\providecommand{\NPA}{Nucl. Phys. A } %
\providecommand{\NPB}{Nucl. Phys. B } %
\providecommand{\PLB}{Phys. Lett. B } %
\providecommand{\PR}{Phys. Rep. } %
\providecommand{\PRL}{Phys. Rev. Lett. } %
\providecommand{\PRC}{Phys. Rev. C } %
\providecommand{\PRD}{Phys. Rev. D } %
\providecommand{\REP}{Phys. Rep. }%
\providecommand{\RPP}{Rep. Prog. Phys. } %
\providecommand{\ZPC}{Z. Phys. C } %
\providecommand{\JHEP}{JHEP } %
\providecommand{\JPG}{J. Phys. G } %

\begin{document}

\begin{frontmatter}

\journal{Nuclear Physics A}

\title{Nuclear EMC Effect in a Statistical Model}

\author{Yunhua Zhang},
\author{Lijing Shao},
\author{Bo-Qiang Ma}\ead{mabq@phy.pku.edu.cn}
\address{School of Physics and State Key Laboratory of Nuclear
Physics and Technology, Peking University, Beijing 100871, China}

\begin{abstract}
A simple statistical model in terms of light-front kinematic
variables is used to explain the nuclear EMC effect in the range $x
\in [0.2,~0.7]$, which was constructed by us previously to calculate
the parton distribution functions (PDFs) of the nucleon. Here, we
treat the temperature $T$ as a parameter of the atomic number $A$,
and get reasonable results in agreement with the experimental data.
Our results show that the larger $A$, the lower $T$ thus the bigger
volume $V$, and these features are consistent with other models.
Moreover, we give the predictions of the quark distribution ratios,
\emph{i.e.}, $q^A(x) / q^D(x)$, $\bar{q}^A(x) / \bar{q}^D(x)$, and
$s^A(x) / s^D(x)$, and also the gluon ratio $g^A(x) / g^D(x)$ for
iron as an example. The predictions are different from those by
other models, thus experiments aiming at measuring the parton ratios
of antiquarks, strange quarks, and gluons can provide a
discrimination of different models.
\end{abstract}

\begin{keyword}
statistical model, EMC effect, parton distribution functions,
nucleon structure

\PACS 12.40.Ee, 13.60.Hb, 24.85.+p, 25.30.Mr
\end{keyword}

\end{frontmatter}

\newpage

\section{Introduction}

Deep inelastic scattering (DIS) process is an efficient tool to
detect the nucleon structure, and the high-energy charged lepton can
also reveal the parton distributions in the nuclear environment. The
nucleons in a nucleus were initially thought to be highly
insensitive to their surroundings, and the only nuclear effect in
DIS was believed to be Fermi motion at large $x$. However, in 1982,
the European Muon Collaboration at CERN discovered that nucleons
inside a nucleus have a remarkably different momentum configuration
as expected, \emph{i.e.}, the structure function for a bound nucleon
differs from that for an isolated one
significantly~\cite{emc83,emc87,b83}, which was named the nuclear
EMC effect (for reviews, see, \emph{e.g.},
Refs.~\cite{bc87,fs88,bp87,a94,gst95,pw00,n03}). The existence of
the unexpected effect opens a window onto the behavior of quarks at
nuclear scale, thus it offers the opportunity to use nuclear targets
for exploring aspects of quark dynamics.

In order to account for the EMC effect, there have been many efforts
and insights implemented in various models, \emph{e.g.}, the cluster
model~\cite{pv81,j83:2,ch83,br90,dpv84,swl92}, the pion excess
model~\cite{j83:2,s83,et83,fpw83,s92,bcw84}, the $x$-rescaling
model~\cite{srw84,hlp86,askv85}, the $Q^2$-rescaling
model~\cite{crr83,j84,c85}, the nucleon swelling
model~\cite{dpv84,hlp86}, and the deconfinement
model~\cite{j83:2,k85}. In some sense, most of these available
models provide a fairly good description, instead of an explanation,
to the phenomena. The cluster model explains the EMC effect by
introducing different ways of packing the constituents into clusters
of 3$n$ ($n \geq 2$) quarks, to mimic the field acting on partons
caused by the nuclear mediate. This was suggested by Pirner \emph{et
al.}~\cite{pv81}, and first applied to the EMC effect by
Jaffe~\cite{j83:2}. Later it was discussed in detail by Carlson and
Havens~\cite{ch83}. Furthermore, Barshay and Rein~\cite{br90}
provided a physical picture for this scenario and predicted specific
$Q^2$ dependence. In the pion excess model, Smith~\cite{s83}
suggested that iron contains $6$$\sim$$12$ more pions which carry
$\sim$5\% of the momentum, and Ericson and Thomas~\cite{et83} showed
that the EMC enhancement in the region $x < 0.3$ can be reproduced
with the excess pions if an attractive force was assumed. Close,
Jaffe \emph{et al.}~\cite{crr83,j84,c85} first discovered that the
difference between the structure functions for a bound nucleon and
that for a free one can be ascribed to the change of the effective
$Q^2$, which was named the $Q^2$-rescaling model, and it was
explained that these distinctions arise as a result of the
difference in the scale of confinement of the nucleon.

On the other hand, due to the complicated non-perturbative effect,
it is still difficult to calculate the parton distribution functions
(PDFs) of the free nucleon absolutely from the first principle
theory of the quantum chromodynamics (QCD) at present. Various
models according to the spirit of QCD have been brought forward,
therein statistical ones, providing intuitive appeal and physical
simplicity, have made amazing
success~\cite{ap82,ap84,cle87,ct88,cdj93,mu89,bha1,bha2,bha3,bha4,bha5,gdr91,dkg94,dm96,sof1,sof2,sof3,sof4,sof5,sof6,sof7,sof8,sof9,sof10,zyj1,zyj2,zyj3,su04,ah05,tmft08,bl90}.
Actually, as can be speculated, with partons bound in the wee volume
of the nucleon, not only the dynamic, but also the statistical
properties, for example, the Pauli exclusion principle, should have
important effect on PDFs. Angelini and Pazzi~\cite{ap82,ap84}, as
pioneers, found that the nucleon valence quark distribution has a
thermodynamical behavior for $x>0.1$ with the temperature decreasing
for different $Q^2$. Cleymans and Thews~\cite{cle87,ct88,cdj93}
explored a statistical way to generate compatible PDFs. Mac and
Ugaz~\cite{mu89} incorporated first order QCD correction, and
afterwards Bhalerao \emph{et al.}~\cite{bha1,bha2,bha3,bha4,bha5}
introduced finite-size correction; they both referred to the
infinite-momentum frame (IMF). Devanathan \emph{et
al.}~\cite{gdr91,dkg94,dm96} proposed a thermodynamical bag model
which evolves as a function of $x$, and the structure functions they
got have correct asymptotic behavior; in addition, they parametrized
on $T$ and exhibited the scaling behavior. Bourrely, Soffer, and
Buccella~\cite{sof1,sof2,sof3,sof4,sof5,sof6,sof7,sof8,sof9,sof10}
developed a new form of statistical parametrization, and by
incorporating QCD evolution they got indeed remarkable PDFs.
Otherwise, Zhang \emph{et al.}~\cite{zyj1,zyj2,zyj3} constructed a
model using the principle of balance without any free parameter, and
the Gottfried sum they got is in surprisingly agreement with
experiments. Later, Singh and Upadhyay~\cite{su04} extended this
model to have spin considered with modifications. Alberg and
Henley~\cite{ah05} tracked the detailed balance model for a hadron
composed of quark and gluon Fock states and obtained parton
distributions for the proton as well as the pion. Recently, Trevisan
\emph{et al.}~\cite{tmft08} presented a statistical model with a
confining potential and took gluon splitting into account further.

The statistical idea is also applied to the nuclear EMC effect.
Angelini and Pazzi~\cite{ap85} introduced thermodynamical analysis
to the EMC effect, and by utilizing the ratio of valence quark
distributions at different temperatures and confinement volumes,
they fit data well. Afterwards, Li and Peng~\cite{lp88} discussed
the EMC effect using the Fermi-Dirac distribution for fermions and
Bose-Einstein distribution for gluons. Further, Ro\.{z}ynek and
Wilk~\cite{rw02} combined nuclear Fermi motion with statistical
effect~\cite{zyj1,zyj2,zyj3} to account for the rise at large $x$.

\begin{figure}
\begin{center}
\includegraphics[scale=0.85]{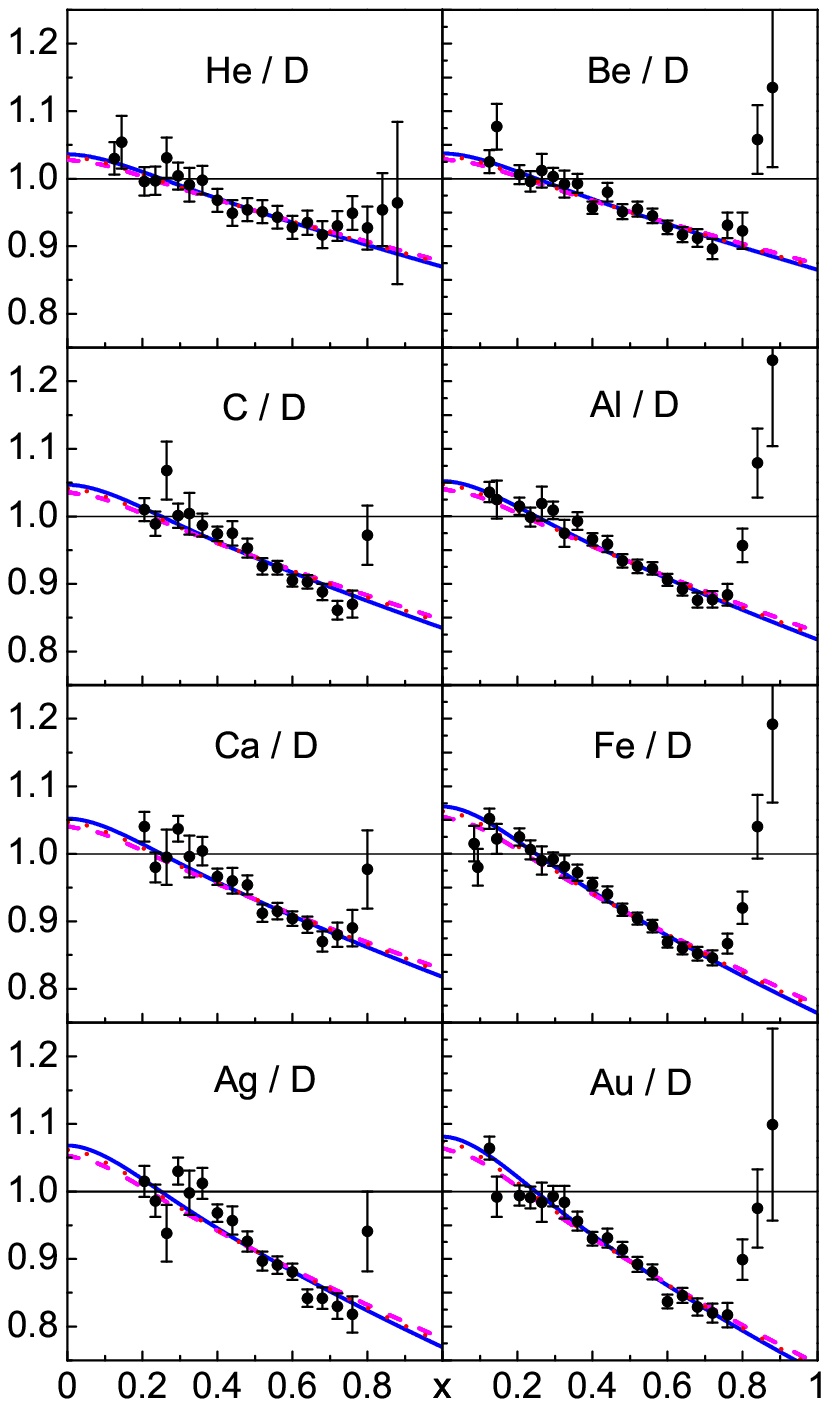}
\renewcommand{\baselinestretch}{2.0}
\caption{The ratios of the structure function $F^A_2/F^D_2$. The
solid, dashed, and dotted curves correspond to three conditions: a)
just considering the $u$, $d$ flavors and the gluon, b) considering
the $u$, $d$, $s$ flavors and the gluon with $m_s=70$ MeV, c)
considering the $u$, $d$, $s$ flavors and the gluon with $m_s=130$
MeV, respectively. The experimental data are from
Ref.~\cite{slac94}.}\label{rf2d}
\end{center}
\end{figure}

In Ref.~\cite{zsm09}, we performed a pure statistical model in terms
of light-front kinematic variables, without any arbitrary parameter
or extra corrected term put by hand, and obtained the analytic PDFs
of free nucleons, \emph{i.e.}, the proton and the neutron. The
results imply that the statistical effect is important to some
aspects of the nucleon structure, and the simpleness and intuition
of the model encourage us to apply it to the EMC effect. Here, we
introduce the temperature $T$ as a parameter versus the atomic
number $A$, and the fit to the experimental data in the region
$x\in[0.2,~0.7]$ is rather reasonable, illustrated in
Fig.~\ref{rf2d}, which indicating that the statistical effect may be
an important source to the EMC phenomenon. But in the high-$x$
region, where Fermi motion dominates to the effect, which is not
considered in our model, our fit is below the data as expected. The
temperature we get is about 1$\sim$2 MeV lower in bound nucleons
than in free ones, and jointly the volume is bigger about
5\%$\sim$10\%. Moreover, we also give the predictions of the quark
distribution ratios, \emph{i.e.}, $q^A(x) / q^D(x)$, $\bar{q}^A(x) /
\bar{q}^D(x)$, and $s^A(x) / s^D(x)$, and the gluon ratio $g^A(x) /
g^D(x)$ for iron as an example.

\section{Statistical approach}
We assume that the nucleon is a thermal system in equilibrium, made
up of free partons (quarks, anti-quarks, and gluons), and the quarks
and anti-quarks satisfy the Fermi-Dirac distribution while the
gluons obey the Bose-Einstein distribution. Instead of boosting the
distribution functions to the
IMF~\cite{mu89,bha1,bha2,bha3,bha4,bha5}, we transform them in terms
of light-front kinematic variables in the nucleon rest frame, and
get $x$-dependent PDFs analytically~\cite{zsm09}
\begin{eqnarray} \label{fx}
f(x) &=&
\pm\frac{g_fMTV}{8\pi^2}\left\{\left(Mx+\frac{m_f^2}{Mx}\right)\,
\ln\left[1\pm
e^{-\frac{\frac{1}{2}\left(Mx+\frac{m_f^2}{Mx}\right)-\mu_f}{T}}
\right]\right.\nonumber\\
&& \left. \qquad -2T\text{Li}_2\left(\mp
e^{-\frac{\frac{1}{2}\left(Mx+\frac{m_f^2}{Mx}\right)-\mu_f}{T}}\right)\right\}\,
\theta\left(x-\frac{m_f^2}{M^2}\right)\,,
\end{eqnarray}
with the upper sign for fermions (quarks, anti-quarks), and nether
sign for bosons (gluons); $g_f$ is the degree of color-spin
degeneracy, which is 6 for quark (anti-quark) and 16 for gluon;
$\mu_f$ is the corresponding chemical potential, while for
anti-quark $\mu_{\bar{q}}=-\mu_q$, and for gluon $\mu_g=0$; $x$ is
the light-front momentum fraction of the nucleon carried by the
specific parton; $M$ is the mass of the nucleon, and the value is
taken as 938.27~MeV; and $\text{Li}_2(z)$ is the polylogarithm
function, defined as $\text{Li}_2(z)=\sum_{k=1}^\infty z^k/k^2$.
Note that the analytic expression above is different from those
attained in the previous statistical
models~\cite{cle87,ct88,cdj93,mu89,bha1,bha2,bha3,bha4,bha5,gdr91,dkg94,dm96}.

The expression of the PDFs we derived is available for all the
partons with mass $m_f$, however, for simplicity, in
Ref.~\cite{zsm09}, we mainly focused on the $u$, $d$ flavors, which
can be viewed massless as a quite good approximation. In this paper,
we will take a further consideration of the $s$ flavor, whose mass
is around 100 MeV, so the constraint $x \geq m_s^2 / M^2$ should be
fulfilled~\cite{z97}. For convenience of a universal manipulation to
all partons, we have multiplied explicitly to the expression of the
PDFs, \emph{i.e.}, Eq.~(\ref{fx}), a factor of the step-function,
$\theta (x - m_f^2 / M^2)$.

In practice, the PDFs in a certain system should be constrained with
some conversation laws. For example, in the proton, they are
\begin{equation}
\label{uv} u_V = \int [u(x)-{\bar{u}}(x)]\,\mathrm{d}x=2\;,
\end{equation}
\begin{equation}
\label{dv} d_V = \int [d(x)-{\bar{d}}(x)]\,\mathrm{d}x=1\;,
\end{equation}
\begin{equation}
\label{xnor} \sum_f\int xf(x)\,\mathrm{d}x=1\;,
\end{equation}
where $f$, in Eq.~(\ref{xnor}), denotes to $u$, $\bar{u}$, $d$, $
\bar{d}$, $g$ if just the $u$, $d$ flavors and the gluon are
considered, and to $u$, $\bar{u}$, $d$, $\bar{d}$, $s$, $ \bar{s}$,
$g$ when the $s$ flavor is appended. However, the contribution of
the heavier flavors, the $c$ quark for instance, is
negligible~\cite{zsm09}.

There are four unknown parameters $T$, $V$, $\mu_u$, $\mu_d$ and
three constraints, thus for a fixed $T$, the rest parameters $V$,
$\mu_u$, $\mu_d$ can be determined uniquely, whereafter we can
obtain all the PDFs. Worthy to mention that, including the $s$
flavor will not introduce any extra parameter, due to
$\mu_{\bar{s}}=\mu_s=0$.

After the PDFs of the proton have been determined, we can get the
PDFs of the neutron as well, using the $p$-$n$ isospin symmetry,
\emph{i.e.}, $u^n(x)=d^p(x)$, $d^n(x)=u^p(x)$,
$\bar{u}^n(x)=\bar{d}^p(x)$, $\bar{d}^n(x)=\bar{u}^p(x)$,
$g^n(x)=g^p(x)$, and $\bar{s}^n(x)=s^n(x)=s^p(x)=\bar{s}^p(x)$.

Eqs.~(\ref{uv})-(\ref{xnor}) should be satisfied for free nucleons,
as well as the ones immersing in the nuclear environment. Here we
mainly assume that a nucleon under a different nuclear circumstance
is equivalent to at a different temperature, and subsequently along
with different $V$, $\mu_u$, and $\mu_d$.

Also note that, the PDFs and the nucleon structure function
\begin{equation}
F_2(x)=x\sum_fe_f^2f(x)\;,
\end{equation}
where $e_f$ is the charge of the parton of flavor $f$, as well as
the Gottfried sum
\begin{equation}
\label{sg}
S_G\equiv\int^{1}_{0}\frac{F^{p}_{2}(x)-F^{n}_{2}(x)}{x}\;\mathrm{d}x
=\frac{1}{3}-\frac{2}{3}\int_0^1\left[\bar{d}(x)-\bar{u}(x)\right]
\mathrm{d}x\;,
\end{equation}
are actually all $T$-dependent.

Afterward, we perform the numerical calculation to determine the
temperatures of the nucleons in different nuclear environment, first
in deuteron, then in other nuclei. Since deuteron is a weakly bound
system, the structure function of it can be taken as a free
isoscalar nucleon approximately, \emph{i.e.},
$F_2^D(x)=\left[F_2^p(x)+F_2^n(x)\right]/2$. In practice, we figure
out $T^D$, \emph{i.e.}, the temperature of nucleon in deuteron by
using $S_G(T^D)=S_G^{exp}$, and then fit $F_2^A(T^A,x)/F_2^D(T^D,x)$
to available experimental data in the region $x \in [0.2,~0.7]$
where the EMC effect dominates, thus the parameters of the nucleon
in the nucleus with atomic number $A$ can be determined.

\section{Results and Discussions}


Numerical calculation is performed under three different conditions
below
\begin{itemize}
\item
just considering the $u$, $d$ flavors and the gluon,
\item
including the $s$ flavor and taking the mass of $s$ quark $m_s=130$
MeV,
\item
including the $s$ flavor and taking $m_s=70$ MeV;
\end{itemize}
here 130 MeV and 70 MeV are respectively the upper and lower limit
of the mass of $s$ quark suggested in PDG2008~\cite{PDG08}.

In practice, as mentioned above, we figure out the temperature of
the nucleon in deuterium $T^D$ firstly, at which the Gottfried sum
$S_G^D$ equals to the experimental value $S_G^{\mathrm{exp}} =
0.235\pm 0.026$~\cite{nmc91,nmc94}. According to the three different
conditions listed above, the results are a bit different
\begin{itemize}
\item
$T^D=$ 46.84 MeV when only $u$, $d$ and $g$ are considered,
\item
$T^D=$ 42.53 MeV when $u$, $d$, $s$ and $g$ are considered with
$m_s=130$ MeV,
\item
$T^D=$ 40.08 MeV when $u$, $d$, $s$ and $g$ are considered with
$m_s=70$ MeV.
\end{itemize}

Then we fit the analytic expression $F_2^A(T^A,x)/F_2^D(T^D,x)$ to
various experimental data from Ref.~\cite{slac94} in the region $x
\in [0.2,~0.7]$ under the constraints aforementioned, and we attain
the corresponding temperatures of the nucleons inside various
nuclei. The fitting curves of some nuclei are shown in
Fig.~\ref{rf2d}. We can see that, when $x$ decreases from 0.2 to
about 0.1, our curves predict the right trend and fit to the
experimental data roughly; but conflicted with the data, as $x$
decreases further from 0.1, our curves still go up and stand as a
constant slightly above unity when $x$ approaches 0. In the high-$x$
direction, the experimental points reveal an apparent rise from
about $x=0.7$$\sim$ 0.8 as $x$ increases, however, our curves depart
from them and monotonously go down, caused by the neglect of Fermi
motion of the nucleons.

As generally recognized now, the fact that $F_2^A\neq F_2^D$ is
referred to with different terms: the behavior of
$F_2^A(x)/F_2^D(x)$ in the region $0.2 \leq x \leq 0.7$ is often
referred to as the ``(special) EMC effect'', where the ratio goes
down straight to lower than unity; in the region $x\leq
0.05$$\sim$0.1, the ratio is smaller than unity, named
``shadowing'', and in the range $0.1 \leq x \leq 0.2$, the ratio
shows an enhancement and becomes a little larger than unity, called
``anti-shadowing''; further, in the region $x \geq 0.7$$ \sim $0.8,
the ratio has a precipitous rise and grows to above unity, where is
the mentioned ``Fermi motion'' region. Here, our model describes the
``EMC effect'' reasonably but fails in other regions, like most of
the other models not considering the effect of ``Fermi motion'' and
``shadowing effect''.

\begin{figure}
\begin{center}
\includegraphics[scale=0.9]{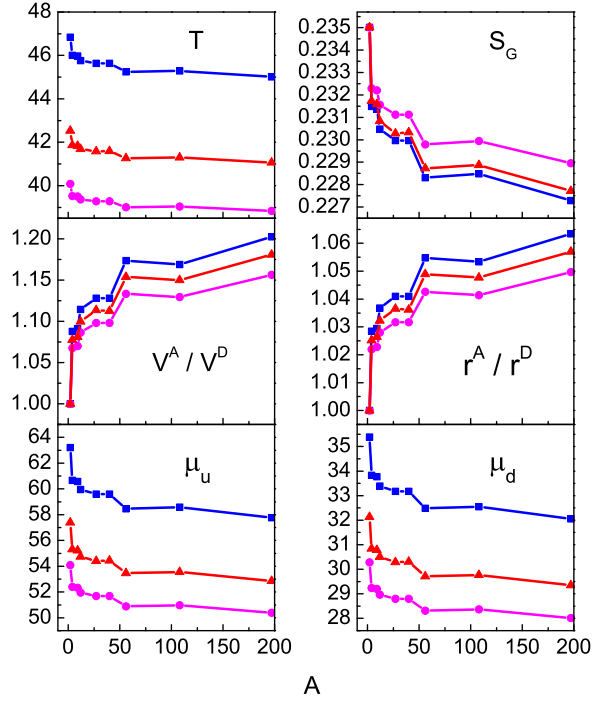}
\renewcommand{\baselinestretch}{2.0}
\caption{The parameters of the nucleons in nuclei --- $T$, $S_G$,
$V^A/V^D$, $r^A/r^D$, $\mu_u$, and $\mu_d$ versus the atomic number
$A$. $T$, $\mu_u$, and $\mu_d$ are in units of MeV. The squared,
circular, and triangular points correspond to the three conditions:
a) just considering the $u$, $d$ flavors and the gluon, b)
considering the $u$, $d$, $s$ flavors and the gluon with $m_s=70$
MeV, c) considering the $u$, $d$, $s$ flavors and the gluon with
$m_s=130$ MeV, respectively.}\label{paraa}
\end{center}
\end{figure}

Once the temperatures of the nuclei are determined, the other
parameters are also fixed simultaneously concerning
Eqs.~(\ref{uv})-(\ref{xnor}), and the values are listed in
Tables~\ref{tab2f},~\ref{tab3fs130},~\ref{tab3fs70} respectively
according to the three different conditions noted above. For an
intuitive view, we present the physical quantities $T$, $S_G$, $V$,
$r$, $\mu_u$, and $\mu_d$ in Fig.~\ref{paraa}, where $r =
(3V/4\pi)^{1/3}$ represents the radius of nucleon from the model. We
can see clearly that the nucleus with larger $A$ has a lower
temperature $T$, jointly a larger volume $V$ and smaller quark
chemical potentials $\mu_u$, $\mu_d$.

The features derived above have an self-consistent explanation in a
physical picture. When $A$ increases, the volume, in which the
partons move, becomes larger, thus the average kinetic energy of the
partons decreases, due to the uncertainty principle. This agrees
with the corresponding lower $T$ property. On the other hand, as $V$
increases, 
the number densities of quarks decrease, thus the chemical
potentials, which are monotonic functions of the corresponding
number densities, become lower.

Our result that the volume of the nucleon in nuclei becomes larger
is qualitatively consistent with other models, such as the
$Q^2$-rescaling model and the nucleon swelling model, though the
starting point of our model is different. The result we derive,
$T^D-T^{\rm Fe} =1$$\sim$2 MeV, is very close to $3\pm 1$ MeV given
in Ref.~\cite{ap85}, and our enhancement in $V^A / V^D$ is about
half of that presented in Ref.~\cite{lp88}. Worthy to note that,
including the $s$ flavor and taking different masses of it lead to
some slightly difference in results, so the $s$ flavor is considered
as a modification here.

\begin{figure}
\begin{center}
\includegraphics[scale=0.9]{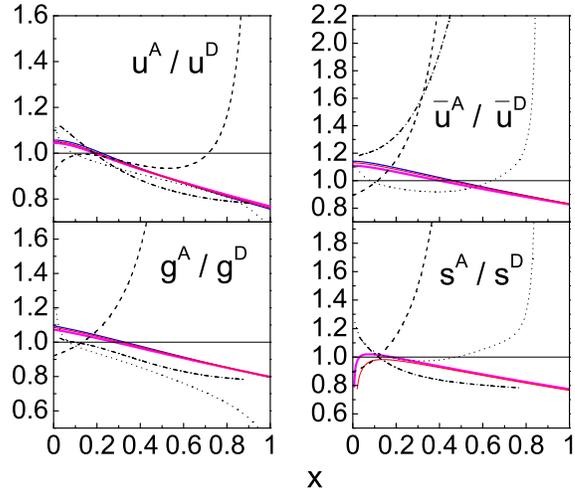}
\renewcommand{\baselinestretch}{2.0}
\caption{The ratios of the PDFs of iron to deuterium. The solid
normal, thick, and thin curves are our results, corresponding
respectively to the three conditions: a) just considering the $u$,
$d$ flavors and the gluon, b) considering the $u$, $d$, $s$ flavors
and the gluon with $m_s=70$ MeV, c) considering the $u$, $d$, $s$
flavors and the gluon with $m_s=130$ MeV. The dashed, dotted, and
dash-dotted curves correspond to the predictions of the cluster
model, the $Q^2$-rescaling model, and the pion excess model,
respectively.}\label{pdfsfe}
\end{center}
\end{figure}

In addition, as parameters are determined, we can give explicitly
the predictions of PDFs of the nucleons inside different nuclei. The
ratios of the PDFs of iron to deuterium are depicted in
Fig.~\ref{pdfsfe} for example. Note that, the distribution of $d$
flavor, not presented in Fig.~\ref{pdfsfe}, is just the same as that
of $u$ flavor, for that all nuclei here, including iron, have been
adjusted to be isospin scalar. The solid normal, thick, and thin
curves correspond respectively to our results in the three
conditions mentioned above, and they almost overlap in each sub
figure, though the parameters of them are a bit different (see
Tables~\ref{tab2f},~\ref{tab3fs130},~\ref{tab3fs70}). The sub figure
of $s^A/s^D$ in Fig.~\ref{pdfsfe} lacks the normal solid curve, for
it is the condition where no $s$ flavor is included.

The ratios in Fig.~\ref{pdfsfe} are only predicted in the range $x
\in [0.2,~0.7]$ since we only fit to the data in this special region
and do not include other effect. We can see that the ratios are
monotonous of $x$, indicating more low-$x$ momentum configuration in
iron than in deuterium, which is in consistency with our intuition
of a lower temperature in iron.

And also, we present the fractions of momentum $x$ and structure
function $F_2$ carried by various partons under the three conditions
respectively in Table~\ref{tabfraction}. From the results, we can
see that as $m_s$ getting smaller, $s$ quarks take up more fractions
of momentum and structure function, and the contributions from which
are somehow considerable. The case with no $s$ quarks can be
considered as a limit for $m_s \to \infty $.

Together with our results for iron, the predicted PDFs of other
models, \emph{i.e.}, the cluster model, the pion excess model, and
the $Q^2$-rescaling model, are also illustrated in
Fig.~\ref{pdfsfe}.

For the cluster model, here we consider $6$-quark cluster only, and
employ the PDFs given in Ref.~\cite{swl92}. Also we assume that the
probability to form a $6$-quark cluster in iron equals to
$0.3$~\cite{ch83}.

Concerning the pion model, we adopt the CTEQ6M
parametrization~\cite{p02} for nucleon and the RMS
parametrization~\cite{s92} for pion, then the PDFs in nuclei
read~\cite{s83,et83,lm06}
\begin{equation}\label{pion}
q_i^A(x) = \int_x^1 \frac{{\rm d} y}{y} f_{\pi}^A(y)
q_i^{\pi}(\frac{x}{y}) + \int_x^1 \frac{{\rm d} y}{y} f_N^A(y)
q_i^N(\frac{x}{y})\;,
\end{equation}
where $q_i^{\pi}(x)$ and $q_i^N(x)$ are respectively the parton
distributions for the free pion and nucleon, and we make use of the
``toy model'' in Ref.~\cite{bcw84} to give the probability
$f_{\pi}(y)$ and $f_N^A(y)$ of finding pionic and normal nucleonic
content respectively in iron.

And as to the $Q^2$-rescaling model, we again take the CTEQ6M
parametrization~\cite{p02} and take $\xi=1.83$~\cite{c85} in the
PDFs below~\cite{crr83,j84,c85}
\begin{equation}\label{q}
q^A(x,Q^2) = q^N(x,\xi (Q^2) Q^2)\;.
\end{equation}

As can be seen in Fig.~\ref{pdfsfe}, the predicted PDFs differ
significantly from model to model. And to distinguish various models
and look into the immanent reason of the EMC effect, we suggest more
experiments to identify the PDFs in nuclei, especially for
anti-quarks, the strange quark, and the gluon. The dimuon yield in
Drell-Yan process~\cite{a90,Garvey:2001yq} can detect the sea
content in nuclei. Ref.~\cite{lm06} also advised that the
semi-inclusive hadron productions in DIS are sensitive to the sea
quark content. Further, the process of charmed quarks production in
DIS can probe the gluon constituent via the photon-gluon fusion
mechanism, and the strange quark content can be detected in
$\Lambda$-$K$ process.

\section{Summary}

In this paper, we utilize a concise statistical model to mimic the
nuclear EMC effect. The statistical model was used previously to
calculate the parton distribution functions (PDFs) of the nucleon in
terms of light-front kinematic variables with no arbitrary parameter
or extra corrected term~\cite{zsm09}. Here, we treat the nucleon
temperature $T$ as a parameter of the atomic number $A$ and find
that the nuclear effect can be explained as a shift of $T$ --- as
$A$ increases, $T$ gets lower and consequently the volume $V$
becomes larger; the larger $A$, the more significant influence.
These features are consistent with other models, for example, the
$Q^2$-rescaling model and the nucleon swelling model.

Further, we present the predictions of the quark distribution
ratios, \emph{i.e.}, $q^A(x) / q^D(x)$, $\bar{q}^A(x) /
\bar{q}^D(x)$, and $s^A(x) / s^D(x)$, and also the gluon ratio
$g^A(x) / g^D(x)$ for iron as an example. These predictions are
rather different from those of other available models. Further
experiments are expected to provide more information of the PDFs in
nuclei, especially for anti-quarks, the strange quark, and the
gluon, then we can test various models better.

Though there is only one free parameter, \emph{i.e.}, the nucleon
temperature $T$, in our model, the descriptions to the experimental
data are reasonable in the EMC domain $x \in [0.2,~0.7]$. We suggest
that the statistical property may be a considerable source to the
nuclear EMC effect, and expect to have more relevant experiments to
test the predictions.

\section*{Acknowledgment}

This work is partially supported by National Natural Science
Foundation of China (No.~10721063) and by the Key Grant Project of
Chinese Ministry of Education (No.~305001). It is also supported by
Hui-Chun Chin and Tsung-Dao Lee Chinese Undergraduate Research
Endowment (Chun-Tsung Endowment) at Peking University, and by
National Fund for Fostering Talents of Basic Science (Nos.~J0630311,
J0730316).

\clearpage
\newpage
{\fontsize{18pt}{\baselineskip}\selectfont List of Figures}
\begin{itemize}
\item Fig.1:~~~The ratios of the structure function $F^A_2/F^D_2$. The
solid, dashed, and dotted curves correspond to three conditions: a)
just considering the $u$, $d$ flavors and the gluon, b) considering
the $u$, $d$, $s$ flavors and the gluon with $m_s=70$ MeV, c)
considering the $u$, $d$, $s$ flavors and the gluon with $m_s=130$
MeV, respectively. The experimental data are from
Ref.~\cite{slac94}.

\item Fig.2:~~~The parameters of the nucleons in nuclei --- $T$, $S_G$,
$V^A/V^D$, $r^A/r^D$, $\mu_u$, and $\mu_d$ versus the atomic number
$A$. $T$, $\mu_u$, and $\mu_d$ are in units of MeV. The squared,
circular, and triangular points correspond to the three conditions:
a) just considering the $u$, $d$ flavors and the gluon, b)
considering the $u$, $d$, $s$ flavors and the gluon with $m_s=70$
MeV, c) considering the $u$, $d$, $s$ flavors and the gluon with
$m_s=130$ MeV, respectively.

\item Fig.3:~~~The ratios of the PDFs of iron to deuterium. The solid
normal, thick, and thin curves are our results, corresponding
respectively to the three conditions: a) just considering the $u$,
$d$ flavors and the gluon, b) considering the $u$, $d$, $s$ flavors
and the gluon with $m_s=70$ MeV, c) considering the $u$, $d$, $s$
flavors and the gluon with $m_s=130$ MeV. The dashed, dotted, and
dash-dotted curves correspond to the predictions of the cluster
model, the $Q^2$-rescaling model, and the pion excess model,
respectively.
\end{itemize}

\clearpage
\newpage
{\fontsize{18pt}{\baselineskip}\selectfont List of Tables}
\begin{itemize}
\item Table 1:~~~Considering the $u$, $d$ flavors and the
gluon.

\item Table 2:~~~Considering the $u$, $d$, $s$ flavors and the gluon
($m_s=130$ MeV).

\item Table 3:~~~Considering the $u$, $d$, $s$ flavors and the gluon
($m_s=70$ MeV).

\item Table 4:~~~The fractions of $x$ and $F_2$ carried by $u$, $\bar{u}$,
$d$, $\bar{d}$, $g$, $s(\bar{s})$ in $^2$D and $^{56}$Fe
respectively under the three conditions: a) just considering the
$u$, $d$ flavors and the gluon, b) considering the $u$, $d$, $s$
flavors and the gluon with $m_s=70$ MeV, c) considering the $u$,
$d$, $s$ flavors and the gluon with $m_s=130$ MeV.
\end{itemize}

\clearpage
\newpage

\clearpage
\newpage

\clearpage
\newpage

\clearpage
\newpage

\begin{table}[Center]
\renewcommand{\baselinestretch}{2.0}
\caption{Considering the $u$, $d$ flavors and the
gluon.}\label{tab2f}\vspace{0.5cm}
\begin{tabular}{c|ccccccccc}\hline
 & $^2$D & $^4$He & $^9$Be & $^{12}$C & $^{27}$Al & $^{40}$Ca &
$^{56}$Fe &
$^{107}$Ag & $^{197}$Au\\
\hline\hline $T~(\text{MeV})$ & 46.84 & 46.00 & 45.97 & 45.76 &
45.64 & 45.64 & 45.25 & 45.29 &
45.01\\
\hline $\mu_u~(\text{MeV})$ & 63.21 & 60.67 & 60.58 & 59.96 & 59.60
& 59.60 & 58.46
& 58.58 & 57.77\\
\hline $\mu_d~(\text{MeV})$ & 35.39 & 33.83 & 33.77 & 33.39 & 33.17
& 33.17 & 32.48& 32.55 &
32.06\\
\hline $V^A/V^D$ & 1.000 & 1.088 & 1.091 & 1.114 & 1.128 & 1.128 &
1.174 & 1.169
& 1.203\\
\hline $r^A/r^D$ & 1.000 & 1.028 & 1.029 & 1.037 & 1.041 & 1.041 &
1.055 & 1.053
& 1.063\\
\hline $S_G$ & 0.235 & 0.232 & 0.231 & 0.231 & 0.230 & 0.230 & 0.228
& 0.229
& 0.227\\
\hline
\end{tabular}
\end{table}

\clearpage
\newpage

\begin{table}[center]
\renewcommand{\baselinestretch}{2.0}
\caption{Considering the $u$, $d$, $s$ flavors and the gluon
($m_s=130$ MeV).}\label{tab3fs130}\vspace{0.5cm}
\begin{tabular}{c|ccccccccc}\hline
 & $^2$D & $^4$He & $^9$Be & $^{12}$C & $^{27}$Al & $^{40}$Ca &
$^{56}$Fe &
$^{107}$Ag & $^{197}$Au\\
\hline\hline $T~(\text{MeV})$ & 42.53 & 41.87 & 41.84 & 41.69 &
41.58 & 41.59 & 41.27 & 41.30 &
41.07\\
\hline $\mu_u~(\text{MeV})$ & 57.39 & 55.31 & 55.21 & 54.75 & 54.41
& 54.44 & 53.46 &
53.55 & 52.86\\
\hline $\mu_d~(\text{MeV})$ & 32.13 & 30.85 & 30.79 & 30.50 & 30.29
& 30.31 & 29.71 & 29.77 &
29.34\\
\hline $V^A/V^D$ & 1.000 & 1.078 & 1.081 & 1.100 & 1.114 & 1.112 &
1.154 & 1.150
& 1.181\\
\hline $r^A/r^D$ & 1.000 & 1.025 & 1.026 & 1.032 & 1.037 & 1.036 &
1.049 & 1.048
& 1.057\\
\hline $S_G$ & 0.235 & 0.232 & 0.232 & 0.231 & 0.230 & 0.230 & 0.229
& 0.229
& 0.228\\
\hline
\end{tabular}
\end{table}

\clearpage
\newpage

\begin{table}[center]
\renewcommand{\baselinestretch}{2.0}
\caption{Considering the $u$, $d$, $s$ flavors and the gluon
($m_s=70$ MeV).}\label{tab3fs70}\vspace{0.5cm}
\begin{tabular}{c|ccccccccc}\hline
 & $^2$D & $^4$He & $^9$Be & $^{12}$C & $^{27}$Al & $^{40}$Ca &
$^{56}$Fe &
$^{107}$Ag & $^{197}$Au\\
\hline\hline $T~(\text{MeV})$ & 40.08 & 39.52 & 39.50 & 39.37 &
39.28 & 39.28 & 39.01 & 39.04 &
38.84\\
\hline $\mu_u~(\text{MeV})$ & 54.08 & 52.39 & 52.33 & 51.95 & 51.68
& 51.68 & 50.88 &
50.97 & 50.39\\
\hline $\mu_d~(\text{MeV})$ & 30.28 & 29.24 & 29.20 & 28.97 & 28.80
& 28.80 & 28.32 & 28.37 &
28.01\\
\hline $V^A/V^D$ & 1.000 & 1.067 & 1.070 & 1.086 & 1.098 & 1.098 &
1.133 & 1.129
& 1.157\\
\hline $r^A/r^D$ & 1.000 & 1.022 & 1.023 & 1.028 & 1.032 & 1.032 &
1.043 & 1.041
& 1.050\\
\hline $S_G$ & 0.235 & 0.232 & 0.232 & 0.232 & 0.231 & 0.231 & 0.230
& 0.230
& 0.229\\
\hline
\end{tabular}
\end{table}

\begin{table}[center]
\renewcommand{\baselinestretch}{2.0}
\caption{The fractions of $x$ and $F_2$ carried by $u$, $\bar{u}$,
$d$, $\bar{d}$, $g$, $s(\bar{s})$ in $^2$D and $^{56}$Fe
respectively under the three conditions: a) just considering the
$u$, $d$ flavors and the gluon, b) considering the $u$, $d$, $s$
flavors and the gluon with $m_s=70$ MeV, c) considering the $u$,
$d$, $s$ flavors and the gluon with $m_s=130$
MeV.}\label{tabfraction}\vspace{0.5cm}
\begin{tabular}{ccc|cccccc}\hline
The fraction & & & $u$ & $\bar{u}$ & $d$ & $\bar{d}$ & $g$ & $s(\bar{s})$\\
\hline\hline & {} & $^2$D & 0.295 & 0.041 & 0.295 & 0.041 & 0.328 &
{}\\
{} & \raisebox{3ex}[0pt]{$x$} & $^{56}$Fe & 0.289 & 0.043 & 0.289 &
0.043 & 0.336 &
{}\\
{} & & $^2$D & 0.703 & 0.097 & 0.176 & 0.024 & {} &
{}\\
\raisebox{8ex}[0pt]{no $s$ quark} & \raisebox{3ex}[0pt]{$F_2$} &
$^{56}$Fe & 0.696 & 0.104 & 0.174 & 0.026 & {} &
{}\\
\hline & {} & $^2$D & 0.253 & 0.035 & 0.253 & 0.035 & 0.281 &
0.072\\
{} & \raisebox{3ex}[0pt]{$x$} & $^{56}$Fe & 0.249 & 0.037 & 0.249 &
0.037 & 0.286 &
0.072\\
{} & & $^2$D & 0.640 & 0.088 & 0.160 & 0.022 & {} &
0.045\\
\raisebox{8ex}[0pt]{$m_s=70~\text{MeV}$} &
\raisebox{3ex}[0pt]{$F_2$} & $^{56}$Fe & 0.634 & 0.093 & 0.158 &
0.023 & {} &
0.046\\
\hline & {} & $^2$D & 0.268 & 0.037 & 0.268 & 0.037 & 0.298 &
0.046\\
{} & \raisebox{3ex}[0pt]{$x$} & $^{56}$Fe & 0.263 & 0.039 & 0.263 &
0.039 & 0.305 &
0.045\\
{} & & $^2$D & 0.664 & 0.091 & 0.166 & 0.023 & {} &
0.028\\
\raisebox{8ex}[0pt]{$m_s=130~\text{MeV}$} &
\raisebox{3ex}[0pt]{$F_2$} & $^{56}$Fe & 0.657 & 0.098 & 0.164 &
0.025 & {} &
0.028\\
\hline
\end{tabular}
\end{table}

\end{document}